# Depletion-Driven Morphological Control of Bundled Actin Networks

James Clarke[a]†, Francis Cavanna[a]†, Anne D. Crowell[b], Lauren Melcher[c], Justin R. Houser[d], Kristin Graham[d], Allison M. Green[b], Jeanne C. Stachowiak[d], Thomas M. Truskett[b], Delia J. Milliron[b], Adrianne M. Rosales[b], Moumita Das[e], and José Alvarado[a]

**The actin cytoskeleton is a semiflexible biopolymer network whose morphology is controlled by a wide range of biochemical and physical factors. Actin is known to undergo a phase transition from a single-filament state to a bundled state by the addition of polyethylene glycol (PEG) molecules in sufficient concentration. While the depletion interaction experienced by these biopolymers is well-known, the effect of changing the molecular weight of the depletant is less well understood. Here, we experimentally identify a phase transition in solutions of actin from networks of filaments to networks of bundles by varying the molecular weight of PEG polymers, while holding the concentration of these PEG polymers constant. We examine the states straddling the phase transition in terms of micro and macroscale properties. We find that the mesh size, bundle diameter, persistence length, and intra-bundle spacing between filaments across the line of criticality do not show significant differences, while the relaxation time, storage modulus, and degree of bundling change between the two states do show significant differences. Our results demonstrate the ability to tune actin network morphology and mechanics by controlling depletant size, a property which could be exploited to develop actin-based materials with switchable rigidity.**

## Introduction

Living cells demonstrate the ability to regulate their internal processes, self-replicate, and participate in higher-order functions in living organisms (1). One of these higher-order functions requires cells to control their outer boundaries. In the case of plant life, most cells are surrounded by a cellulose-based cell wall which confers plants with their shape and stiffness (2). Animal cells are comparatively softer, more dynamic(3), and are able to self-deform in response to external events or to tune their mechanical properties to better accomplish their functions (4). This is partially accomplished by rearrangement and remodeling of the cellular cortex(5), which is a network of biopolymers and molecular motors.

Actin is one of the most prevalent biopolymers in the cell and a key component of the cellular cortex. It exhibits a wide polymorphism in its ability to polymerize and then form higher-order structures out of the polymerized actin (6). Of specific interest for the work presented herein are actin bundles, which are utilized in lamellipodia(7), filopodia(6), stress fibers(8), and microvilli (9). The mechanisms driving bundle formation are varied, and several have been identified in vitro. Electrostatic bundling via counterion condensation screens charges held on the actin, and by electric attraction pulls actin filaments into bundles (10–12). Physiological bundling depends on specific proteins that are known to bind to and crosslink actin(13), such as fascin(14), α-actinin(15), and filamin (16,17).

a. UT Austin Department of Physics, 2515 Speedway, Austin, Texas, USA.
b. UT Austin McKetta Department of Chemical Engineering, E 24th St, Austin, TX 78712.
c. School of Mathematical Sciences, Rochester Institute of Technology, Rochester, NY, USA
d. Department of Biomolecular Engineering, The University of Texas at Austin, Austin, TX,USA
e. School of Physics and Astronomy, Rochester Institute of Technology, Rochester, NY, USA
† These authors contributed equally.

Bundle formation can also be caused by a depletion interaction. Unlike counterion condensation, it is thought that depletion interactions are relevant in living cells (18). In general, depletion interactions can be triggered in a wide variety of contexts: colloids(19), actively stirred rods and beads(20), actin and DNA polymers in solutions(21), DNA polymers suspended with increasing sizes of dextran molecules(22), and microtubules driven by PEG (23). In particular, dilute actin solutions are known to experience self-bundling interactions in systems with sufficient concentration of PEG (24). With large concentrations of PEG, high concentrations of actin are also known to bundle. In this study we elect to use PEG as it is known to bundle actin via a depletion interaction and has an easily tunable molecular weight (25,26). Actin filaments are long compared to the PEG molecules. The persistence length of actin is ~10 μm and the diameter of F-actin is about 7 to 8 nm (27,28), whereas the radius of gyration for PEG, $R_g \sim MW^{3/5}$ (29), is much smaller for the sizes incorporated in this study. For PEG 20kDa, $R_g \simeq 3.9$ nm, and for PEG 6kDa, $R_g \simeq 1.9$ nm (29). Given this separation in length scales, there is an entropic favorability for PEG in sufficient concentration to occupy maximal volume. PEG accomplishes this by depleting the distances between actin filaments in its local environment, thus inducing bundle formation. Osmotically, when two filaments are closer than twice the radius of a PEG molecule, the filaments act as a semi-permeable structure where fluid is permitted to flow, but PEG particles are excluded from the region. This attracts the two filaments toward one another, effectively depleting the separation between them (30).

While a transition in actin network structure has been identified across a sufficiently large concentration of PEG, much less is known about the effect of molecular weight of the depletant with regards to actin bundle structure. This is surprising, because PEG molecular weight is an important control parameter for the depletion interaction (26,31,32). Furthermore, addressing this gap could lead to the development of responsive actin-based materials fueled by control over the molecular weight of the PEG depletants, which can be accomplished by the association or dissociation of individual PEG molecules.

Here we study the effect of PEG molecular weight on the structure of actin networks. In particular, structural measures such as the tendency to form bundles, the bundle diameter, the spacing between bundles, the relaxation time of the network, persistence length, bending modulus, and the bulk mechanical properties of the network are investigated.

## Results

### Microstructure of Bundled Actin Networks.

**Confocal Microscopy.** The primary results of this segment are captured in Figure 1. We observe a phase transition between two morphologically distinct regions of actin networks by systematically varying the molecular weight and concentration of PEG molecules in the assay. In the confocal micrographs of fluorescently tagged actin filaments (panel a) one can observe a distinction between weakly bundled and strongly bundled networks to the left and right of the blue line in Figure 1, respectively. In the weakly bundled regime, coexistence of actin bundles and single-filament actin networks are observed, whereas for strongly bundled samples a network of actin bundles is formed. To quantify the regions between weakly bundled and strongly bundled networks, we developed an image processing algorithm that quantifies the presence of bundles and hence the observed phase transition (see Methods). We find that strongly bundled networks occur for a degree of bundling greater than $\sim 0.2$ [a.u.] (panel b) associated with the observed phase transition.

We are particularly interested in the transition between weakly bundled and strongly bundled states as we vary PEG molecular weight. In order to better characterize this transition, we focus here on two points in phase space, $\mathbf{\Gamma_{6k}}$ and $\mathbf{\Gamma_{20k}}$, which consist of solutions with [actin] = 12 μM, and [PEG] = 0.1% (w/v) at molecular weights of 6kDa and 20kDa, respectively. First, we characterize the mesh size, $\mathbf{\xi}$, of these two states. The results of the mesh size analysis are given in Figure 2.

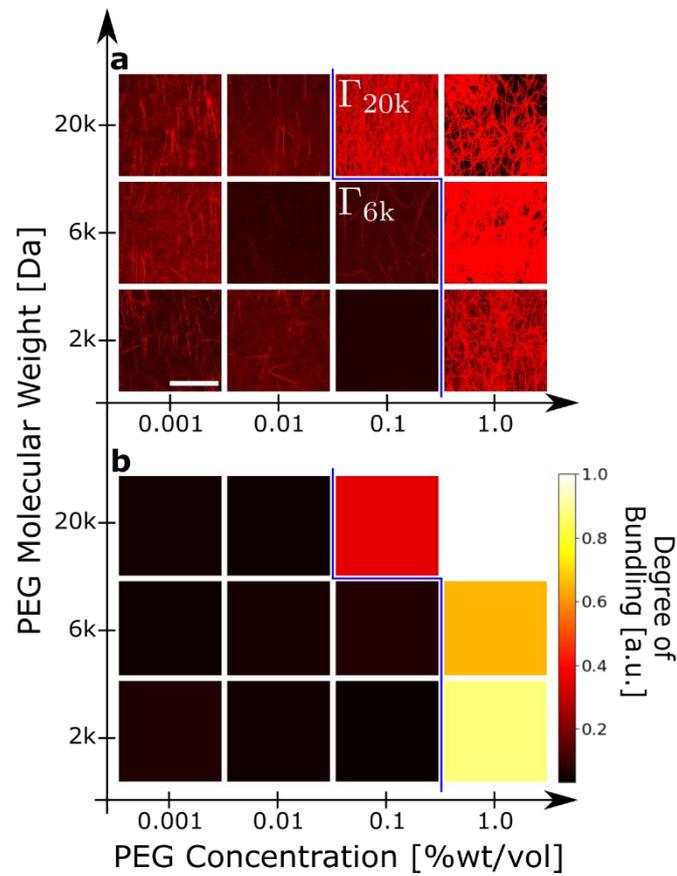

Figure 1: (a) A representative phase diagram of [actin] = 12 μM with various concentrations of PEG on the x-axis and the molecular weight of the PEG on the y-axis corresponding to a total of 36 distinct samples. Scale bar = 40 microns. The critical concentration of bundling (blue line) for PEG 6k and PEG 2k is at [PEG] = 1.0% (w/v), and for PEG 20k at [PEG] = 0.1% (w/v). $\Gamma_{6k}$ & $\Gamma_{20k}$ denote regions of interest for further characterization. (b) Heat maps representing the degree of bundling of the phase space depicted in (a), as found by the skeletonizing algorithm. Each value corresponds to N=3 measurements per sample condition.

Figure 2 shows that the mesh size distributions for the $\Gamma_{6k}$ and $\Gamma_{20k}$ states are statistically indistinguishable from one another. The t-test for this system is close to a significant p-value, but the large spread in the $\Gamma_{6k}$ mesh size results in too high a p-value to give significance. This spread and its impact on proximity to significant separation are within our expectations for this system as these samples are near a morphological phase transition between weakly and strongly bundled states and are more sensitive to sample-to-sample variation which manifests here in the mesh size distribution for $\Gamma_{6k}$.

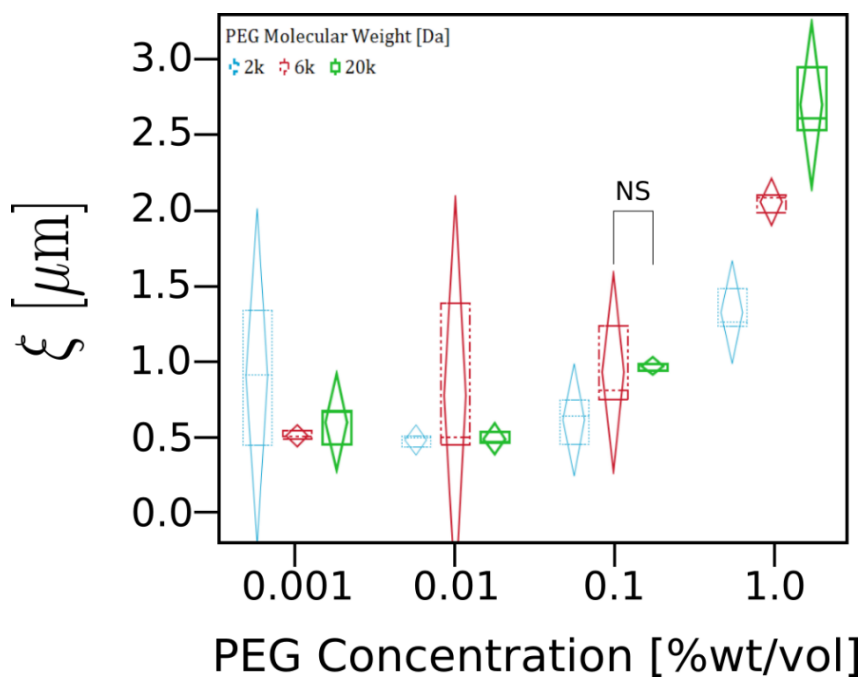

Figure 2: Quartile box plots with confidence diamonds of each network's mean mesh size in microns extracted from confocal micrographs for various concentrations of PEG (N=3 per each sample condition). The molecular weight of the PEG is given in the inset legend where blue fine-dashes, red irregular-dashes, and green solid-line correspond to PEG at molecular weights of 2k, 6k, and 20k, respectively. Direct comparison is drawn between $\Gamma_{6k}$ & $\Gamma_{20k}$ where we can see that the mesh size is statistically identical for the two populations. "NS" denotes that the two populations are statistically indistinguishable.

**Diffusion of actin polymers in Dynamic Light Scattering**

To determine the diffusive properties of $\Gamma_{20k}$, $\Gamma_{6k}$, and a control without PEG, the samples were probed with a Zetasizer Nano ZS instrument, with a q-vector of $|q| = \frac{4\pi \cdot 1.334}{632\,nm} \sin(173°/2) = 26.4\ \mu m^{-1}$ (See Methods). Figure 3 represents a set of correlation curves describing the timescale associated with decorrelation of the system from an original scattering state. Correlation curves for each condition measured with dynamic light scattering are represented.

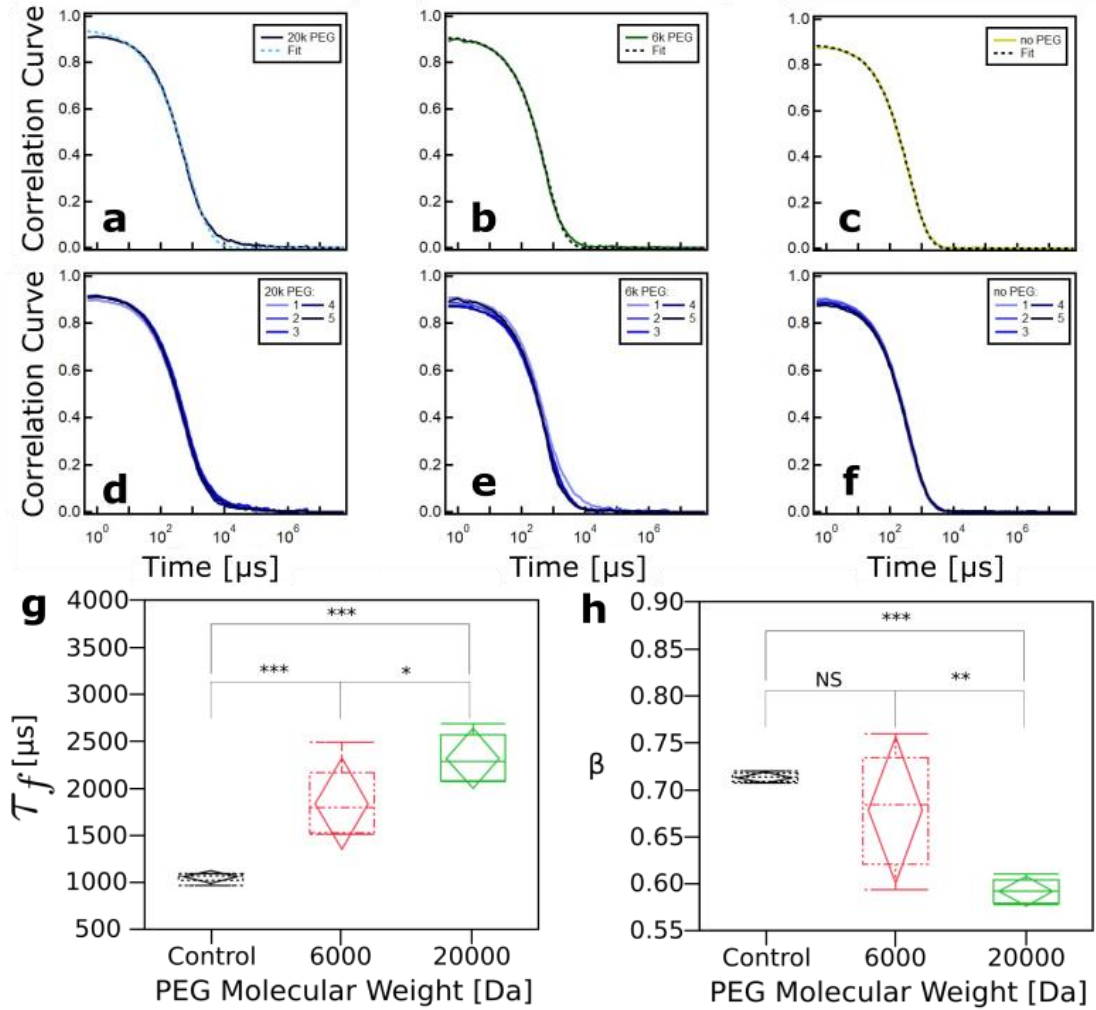

Figure 3: Fitting of correlation function data from dynamic light scattering (DLS) measurements, against $\Gamma_{20k}$ and $\Gamma_{6k}$ and a control of [actin] = 12 μM with no PEG. Panels a-c) contains a sample fitting of one of the five replicates for each condition. Panels d-f) represent the total intensity correlation functions measured for each experimental condition. Each curve was fitted, and a relaxation time and stretch exponent were pulled from the fit. The relaxation time for each condition is represented in g), and was $\tau_{f,20} = 2313.7 \pm 231.0\ \mu s$, $\tau_{f,6} = 1833.1 \pm 351.5\ \mu s$, and $\tau_f = 1052.8 \pm 47.4\ \mu s$. The relaxation time for the $\Gamma_{20k}$ state was significantly different from the relaxation time for the $\Gamma_{6k}$ state (p = 0.0166), and the $\Gamma_{6k}$ state was significantly different from the state with no PEG, (p = 0.0007). The stretch exponent is represented in h), and is $\beta_{20} = 0.593 \pm 0.012$, $\beta_6 = 0.679 \pm 0.056$, and $\beta_c = 0.713 \pm 0.004$. The difference between relaxation times for the control and the $\Gamma_{6k}$ state is not significant (p = 0.1668), but the difference between the $\Gamma_{6k}$ state and the $\Gamma_{20k}$ state is significant (p = 0.0030).

These curves are fit using Eq. 3 (see Methods) to extract relevant parameters. Actin-PEG systems polymerize on the order of tens of seconds, in contrast to systems which experience gelation over dozens of minutes (33,34). Consequently, we observe no polymerization phase, and the plots represent the system at or close to a stabilized state. The relaxation time of the $\Gamma_{20k}$ state, $\tau_{f,20} = 2314 \pm 40\ \mu s$, is roughly twice that of the control state, $\tau_f = 1053 \pm 3\ \mu s$, with the $\Gamma_{6k}$ state somewhere in between, at $\tau_{f,6} = 1833 \pm 22\ \mu s$. Relaxation time corresponds to the time needed for the Intensity correlation function to decay. Longer times correspond to less diffusive networks, and shorter times correspond to networks undergoing more motion. In addition, fitting the correlation function yields the stretching exponent β, which indicates how broad the relaxation times are. β is defined between 0 and 1, and β = 1 corresponds to a single relaxation time, while β < 1 corresponds to an increased broadening of relaxation times as β decreases to 0. The stretched exponential fits reveal the $\Gamma_{20k}$ state has a lower stretching exponent β than that of $\Gamma_{6k}$ and the control, which indicates a more heterogenous distribution for the $\Gamma_{20k}$ than the $\Gamma_{6k}$ states.

The DLS analysis can be extended by considering a dynamic structure factor representative of semiflexible filaments (35–37) given by Eq. 4 (see Methods). This structure factor relates the relaxation rate to the persistence length of the polymers. The results of this are given in Figure 4. We report $L_p^{r6k} = 46\ \mu m$ and $L_p^{r20k} = 66\ \mu m$, given as mean values for the respective state. While we observe shifted means, the distributions on persistence lengths for the $\Gamma_{6k}$ and $\Gamma_{20k}$ are statistically indistinguishable from one another. The actin-only control reported value for persistence length is $L_p^{Control} = 10\ \mu m$ which was tuned to match known values (see Methods) (38).

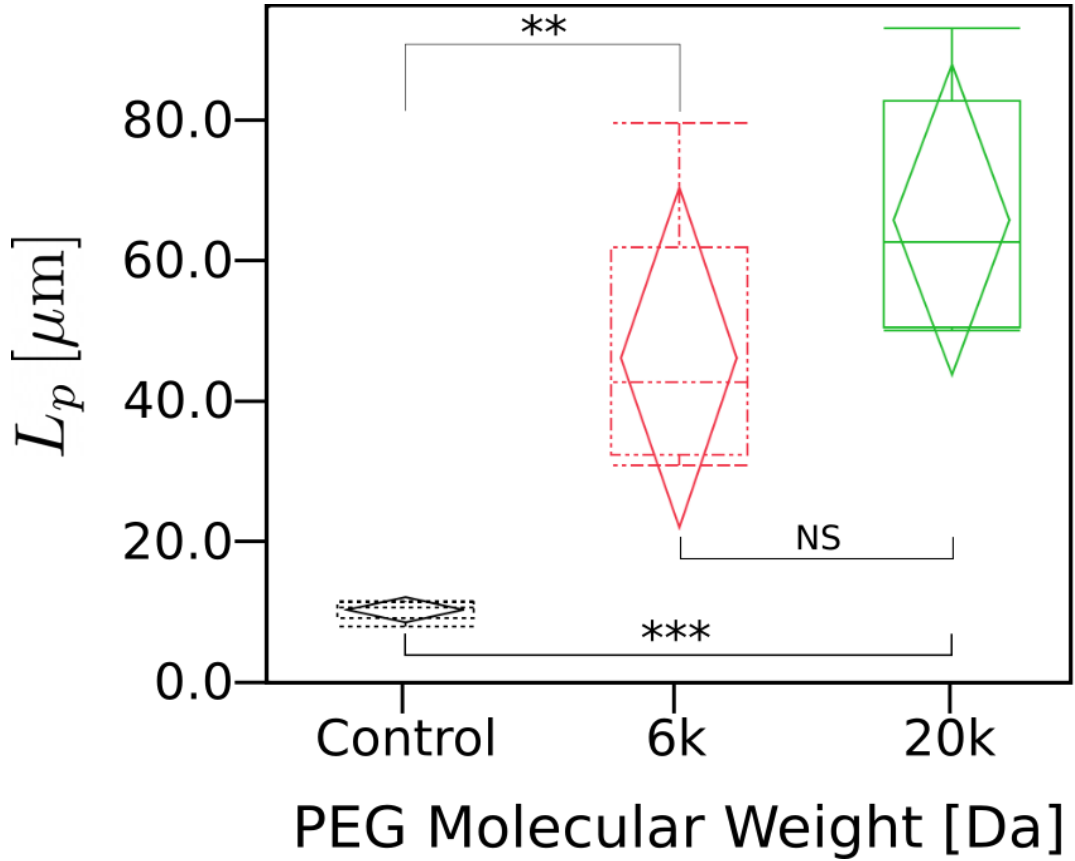

Figure 4: Distributions of measured persistence lengths, $L_p$ for dynamic light scattering experiments. Control corresponds to a filamentous actin meshwork in the absence of PEG molecules. The $\Gamma_{6k}$ and $\Gamma_{20k}$ samples are statistically indistinguishable, per this analysis. This is denoted by "NS". Both $\Gamma_{6k}$ and $\Gamma_{20k}$ are statistically differing from the filamentous actin meshwork.

**Electron Microscopy.** In order to characterize the bundle microstructure of our candidate morphologies, we turn to electron microscopy to deliver better spatial resolution than obtained in the diffraction limited confocal micrographs. Cross-sectional electron micrographs were obtained from cut faces freshly prepared by Ga+ Focused Ion Beam (FIB). The diameters of exposed actin bundles were then measured via Scikit-Image morphological active contours method based on the Chan-Vese algorithm (39). The results of these experiments are presented in Figure 5.

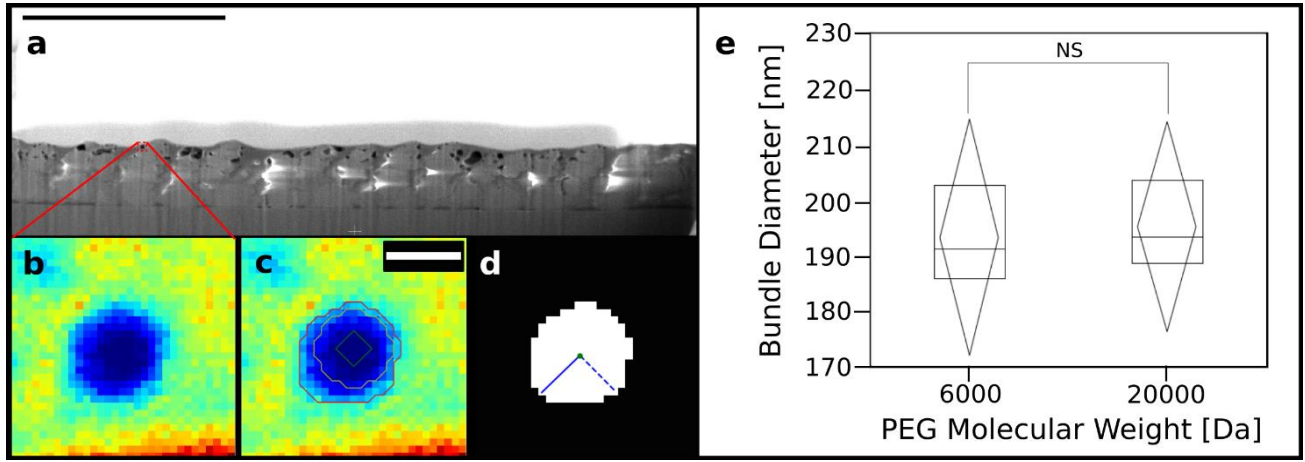

Figure 5: (a) Cross-sectional SEM micrograph of PEG bundled actin network for $\Gamma_{6k}$. Red dashed box indicates the subset of the image corresponding to panels b-d. Scale bar = 5 µm (b) Pre-processed subset of SEM micrograph with one actin bundle. (c) Active contour overlay of SEM micrograph subset seen in panel b. Green, yellow, red represent the first, 5th, and final contours, respectively. Top right white colored scale bar = 100 nm. (d) Resultant area considered for critical dimension measurement of bundle diameter with the centroid (green dot), major axis (solid blue line) & minor axis (dashed blue line) displayed. (e) Quartile box plots with confidence diamonds of mean bundle diameter size in nanometers of network extracted from cross-sectional electron micrographs for the $\Gamma_{6k}$ and $\Gamma_{20k}$ conditions. The two populations are statistically indistinguishable, denoted by "NS" in the panel.

The electron microscopy results indicate that the bundle diameters of $\Gamma_{6k}$ and $\Gamma_{20k}$ are statistically identical. This result suggests that the changing the molecular weight of PEG doesn't change the bundle diameter, as observed by our electron microscopy technique. This result is in agreement with what we see in the confocal micrographs of Figure 1, where PEG bundle sizes visibly change as a function of concentration – yet remain unchanged along the molecular weight axis. The confocal system is diffraction-limited – with aberration-free minimum resolution at 190 nm which coincides with the observed diameters via the scanning electron microscope (SEM).

**Intra-bundle Distance using FRET.**

The $\Gamma_{6k}$ and $\Gamma_{20k}$ states were also characterized using Förster Resonant Energy Transfer (FRET). FRET has been proposed as a sensing mechanism in a myriad of biochemical contexts(40) and utilized to study membrane crowding and steric pressure on membranes (41,42). In this study we utilized FRET as a sensor to measure the distribution in intra-bundle distances between neighboring actin filaments in our system, where perturbations in the fluorescent lifetime can be correlated to real-space distances between a donor-acceptor pair (43). The main results of these experiments are given in Figure 6.

Using the fluorescence lifetimes measured in both $\Gamma_{6k}$ and $\Gamma_{20k}$, $\tau_{DA}$, along with the fluorescence lifetime of the sample without any acceptor fluorophores, $\tau_D$, we calculate the FRET efficiency $E$ with,

$$E = 1 - \frac{\tau_{DA}}{\tau_D}. \tag{1}$$

Given that this is measured on a per-pixel basis, with spatial separation of 300nm per pixel in the sample, the measured efficiency is built using the mean of the respective fluorescence lifetimes, giving a spatially averaged efficiency, $< E >$. With this, the average donor-acceptor distance is calculated using,

$$R_{DA} = R_0 \left( \frac{1}{<E>} - 1 \right)^{\frac{1}{6}}, \tag{2}$$

Where $R_0$ is the Förster radius determined to be 6.72 nm by the choice of donor and acceptor in the system (43).

The main result of the FRET analysis is that the intra-bundle spacing between filaments is statistically indistinguishable across our samples as demonstrated in Figure 6 (44).

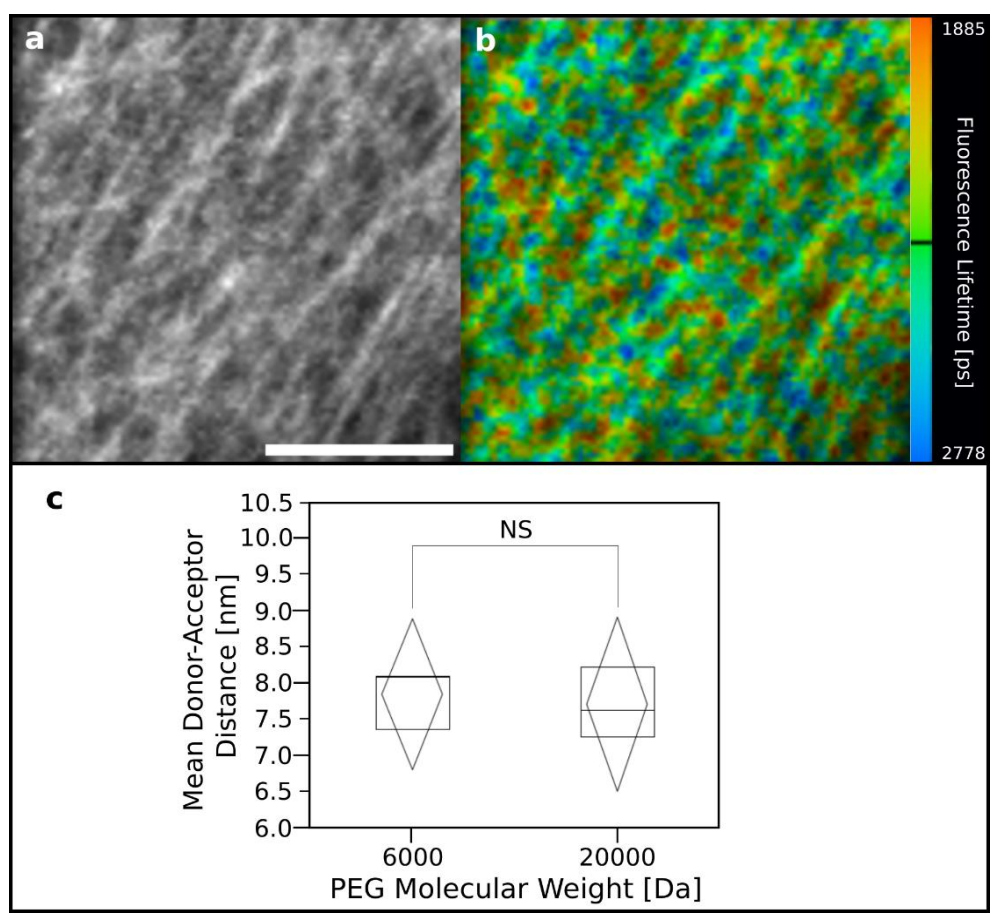

Figure 6: (a) FRET intensity confocal micrograph of a $\Gamma_{6k}$ bundled actin sample. Scale bar = 50 µm  (b) FRET fluorescence lifetime overlay of micrograph shown in a. Colorbar represents the distribution in fluorescence lifetimes measured in the sample. The solid black line represents the mean of the distribution. (c) Quartile box plots and confidence diamonds for mean donor-acceptor distance giving the mean intra-bundle spacing for $\Gamma_{6k}$ and $\Gamma_{20k}$, "NS" denotes that the two distributions are statistically equivalent.

**Bulk Properties of Bundled Actin Networks.** Previous studies characterized the bulk rheological properties of actin networks bundled via physical crosslinks, such as scruin (45,46). The rheological properties of actin networks bundled via depletion forces have also been studied with respect to changes in PEG concentration (25). As previously noted, however, the effect of PEG molecular weight on the bulk rheological properties of bundled actin networks has not been quantified. Thus, in this study, oscillatory shear rheometry is used to characterize the bulk properties of the bundled actin networks. Specifically, we measure the storage modulus ($G'$) and loss modulus ($G''$) as a function of time for the $\Gamma_{6k}$ and $\Gamma_{20k}$ states (Figure 6). $G'$ is especially of interest because it provides a measure of the bulk elasticity of the networks. Three replicates of each network type were created in pairs: for each replicate, enough actin for two networks was prepared, and from this actin preparation, one $\Gamma_{6k}$ sample and one $\Gamma_{20k}$ sample were each prepared. Figure 7a shows the evolution of $G'$ for one pair of replicates. For both networks, $G'$ increases during the first 10 minutes. Because the components of the network are mixed immediately before beginning measurements on the rheometer, the increase in $G'$ during that period is attributed to the formation of the bundled actin networks, corresponding to the gelation time scale (34). After this time, $G'$ plateaus to a steady value, suggesting that network formation is complete. Notably, $G'$ is higher for $\Gamma_{20k}$ than for $\Gamma_{6k}$, suggesting that a stiffer network is formed when a higher molecular weight PEG chain is employed. In each set of replicates, this trend holds (shown in Figure S1). When comparing the equilibrium storage modulus distributions across replicates, a statistically significant difference between $\Gamma_{6k}$ and $\Gamma_{20k}$ is not observed. However, normalizing each replicate to the equilibrium value of $\Gamma_{6k}$ gives significant variation across replicates, as demonstrated in Figure 7b. This normalization procedure eliminates systematic variation associated with pipetting variance between replicates. We find that $\Gamma_{20k}$ replicates are 1.5 times stiffer than $\Gamma_{6k}$ replicates.

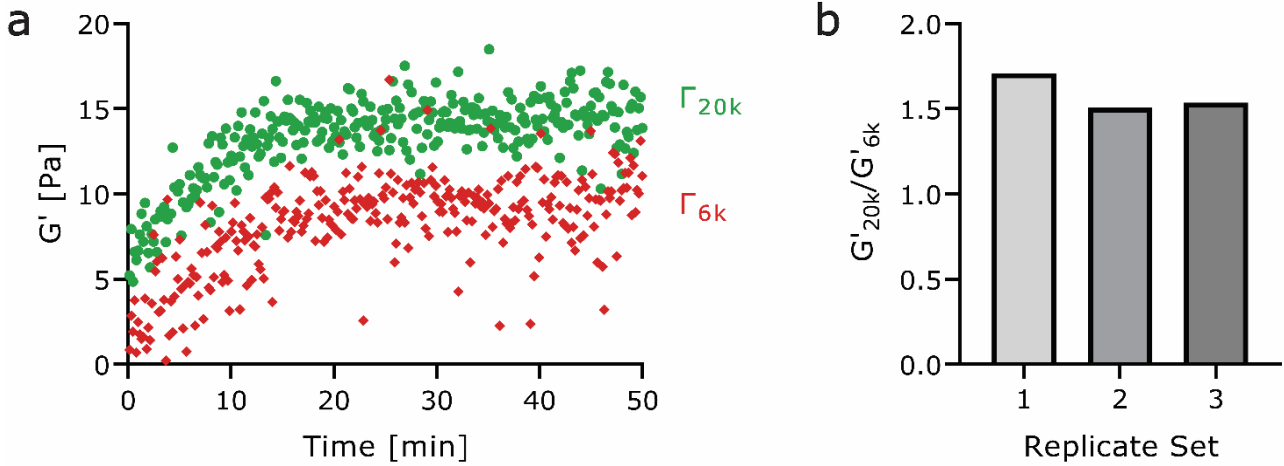

Figure 7: (a) Time sweep showing the evolution of the storage moduli of the bundled actin networks formulated with $\Gamma_{6k}$ (red diamonds) and $\Gamma_{20k}$ (green circles). The gray dashed line shows the minimum modulus that can be measured using the 8 mm parallel plate geometry on the Discovery HR 20 rheometer. (b) Bar graph of normalized storage moduli for all replicates. One-way t-test gives a significant difference between $G'_{20k}/G'_{6k}$ and unity ($G'_{6k}/G'_{6k}$) with a two-way p-value of 0.0116.

**Model and Method.** So far we have presented experimental results demonstrating that PEG concentration and molecular weight sensitively tune the morphology of actin networks (cf. Fig. 1). In order to understand the underlying depletion mechanism, we compare our experimental results to a numerical model, which consists of actin filaments interacting with PEG particles in three dimensions. Each actin filament is modeled as a chain made of beads, and the relative cost of stretching and bending the chain is informed by known mechanical properties of actin filaments. We model the pairwise interactions between any two particles, whether belonging to actin bead-chains, or PEG, or between actin and PEG using a Lennard Jones potential, $V_{LJ} = 4\epsilon \left[ 12 \frac{\sigma_{ab}^{12}}{r_{ij}^{13}} - 6 \frac{\sigma_{ab}^6}{r_{ij}^7} \right]$, where $\sigma_{ab} = \frac{\sigma_a + \sigma_b}{2}$, where $\sigma_a$ and $\sigma_b$ represent the diameters of two reference beads. The interparticle distance between particles with indices $i$ and $j$ is denoted by $r_{ij}$ and the strength of the interaction's potential is defined by $\epsilon$. The interaction potential is truncated to incorporate only repulsive interactions representing a hard core for actin-PEG interactions, while for PEG-PEG and actin-actin interactions we additionally allow attractive interactions to mimic the Van-der Waals interactions.

The motion of the PEG particles follows the overdamped Langevin equation, $\frac{dr}{dt} = \frac{D}{K_b T} F_{LJ} \sqrt{2D} \eta$, where the interparticle interaction force $F_{LJ}$ is derived from the Lennard Jones potential defined above, and the thermal diffusion is represented by the noise term $\sqrt{2D} \eta$, where $D$ is the diffusion constant and $\eta$ is a vector drawn from a Gaussian distribution with a mean of 0 and variance of 1; $K_b$ and T are the Boltzmann constant and room temperature. The motion of an actin bead in strand $i$ is also given by an overdamped Langevin equation: $\frac{dr_i}{dt} = \frac{D}{K_b T} \left( \sum F_{LJ_{ij}} + \sum F_{LJ_{ik}} + \sum F_{LJ_{il}} \right) + \frac{D}{K_b T} \left( \sum F_{s,ij} + \sum F_{b,ij} \right) + \sqrt{2D} \eta$, where $F_{LJ_{ij}}$, $F_{LJ_{ik}}$, and $F_{LJ_{il}}$ are the Lennard Jones interparticle forces, and indices $i$ and $j$ describe intra-strand pairwise interactions, $i$ and $k$ describe inter-strand interactions, and $i$ and $l$ describe interactions between actin and PEG particles. The actin filaments resist stretching (or compression) and bending with forces $F_{s,ij}$ and $F_{b,ij}$, respectively, obtained from the corresponding stretching and bending deformation energies, $U_s = \frac{K_s}{2}(l - l_0)^2$ and $U_b = \frac{K_b}{2}(\theta - \theta_0)^2$. In the equations above, $l$ represents the distance between two nearest-neighbor actin beads in a strand, and $l_0$ describes the equilibrium rest length for actin beads which is given by the actin bead diameter. The stretching stiffness is $K_s$, the bending rigidity is $K_b$ and $\theta$ is the angle between three sequential actin beads and the equilibrium value of this angle, $\theta_0$, is set to $\pi$.

We solve both Langevin equations using the Forward Euler-Maruyama approach. The timestep $dt$ is set to $10^{-3}$ and each simulation was run up to $500\tau$, so the system had enough time to equilibrate and reach a steady state. Simulation results are shown below in Figure 8a along with the representative degree of bundling results as in the confocal analysis. We find qualitative agreement between the phase diagram produced by the simulation and that of experiment. In both simulation and experiment, the actin network exhibits increased bundling as either PEG concentration or PEG molecular weight are increased. However, the transition to a purely bundled state appears to be sharper in the experiment than in simulation (see Discussion).

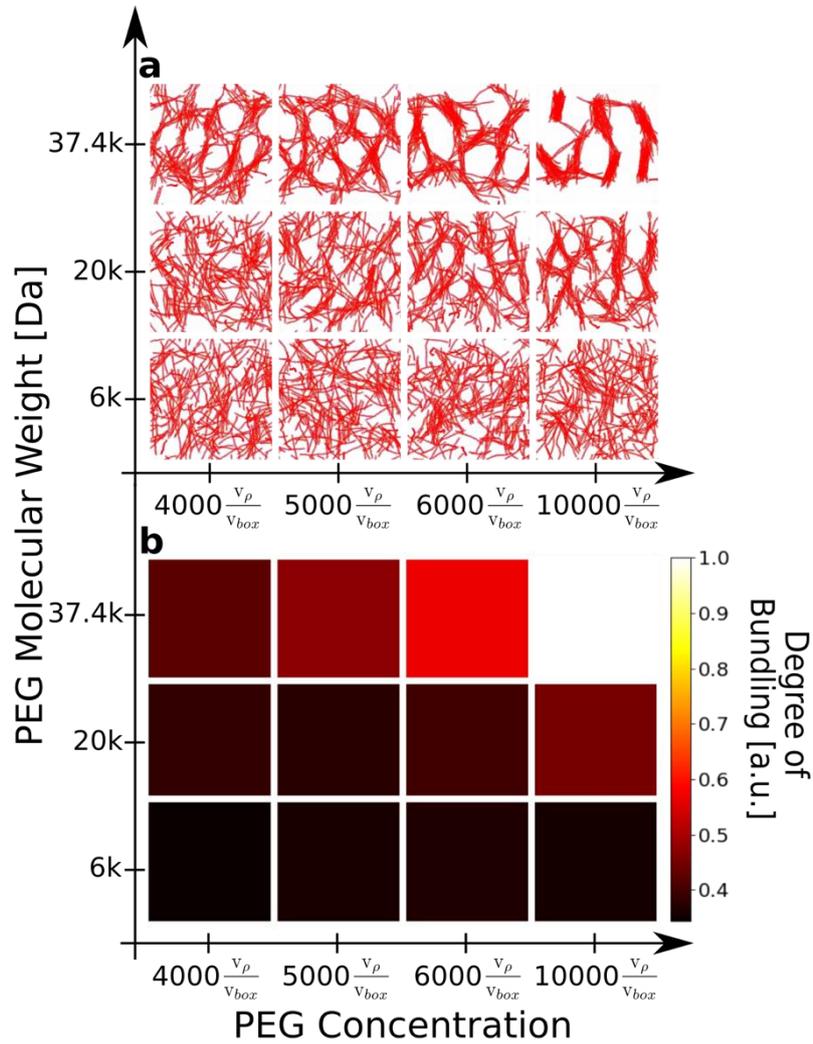

Figure 8: (a) The figures show the simulated equilibrium morphologies of the networks and bundles of actin filaments in the presence of PEG, and demonstrates the role of PEG molecular weight and concentration on the actin filament bundling. As the molecular weight increases (y-axis), significant bundling of actin filaments is observed. Similarly, increasing the concentration of PEG (x-axis) also leads to appreciably more pronounced bundles of actin once the molecular PEG weight is greater than 6k. (b) Degree of bundling results show that the system moves toward more bundled morphologies when molecular weight and concentration of PEG molecules are increased. $v_\rho$ and $v_{box}$ are described in the methods.

## Discussion

The main results of this work are represented in Figure 9.

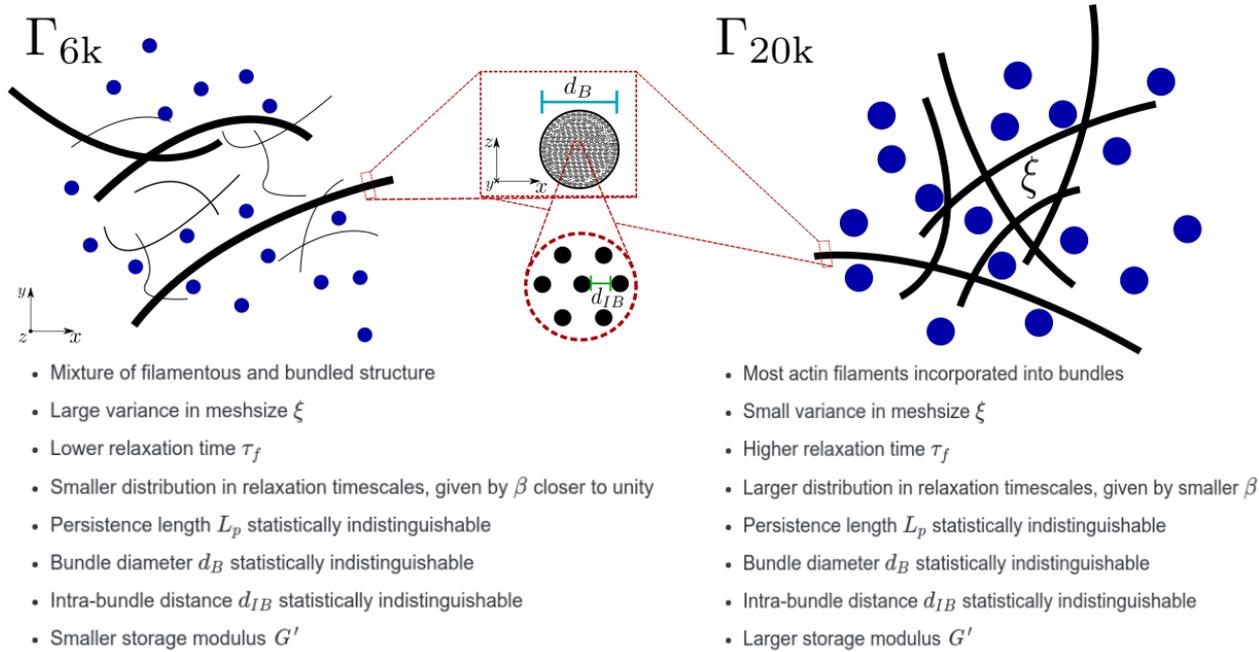

Figure 9: Graphical representation summarizing the main findings of this study. Left: $\Gamma_{6k}$ showing a mixed state of filamentous actin, PEG molecules, and bundled actin with cascading magnified regions showing the measured bundle diameter, $d_B$, and intra-bundle spacing, $d_{IB}$. Right: $\Gamma_{20k}$ state showing bundled actin and PEG molecules. Note: PEG molecules are represented here as spheres whose radius is the PEG molecules radius of gyration $R_g$. The relevant findings from this work are presented in bulleted form below each state.

An increase in PEG MW gives an effective increase in $R_g$ and thus changes the preferred entropic configuration of the system. The depletion interaction is driven by the difference in length scales between system components which give different preferred entropic configurations. The stronger dependence of bundle size, as seen by confocal and electron microscopy, supports the notion that the more important factor for increasing the size of a bundle is the local density of depletion agents that surround a group of filaments, whereas the molecular weight governs how many of the filaments must be depleted for the system to reach an entropically favorable configuration.

Confocal microscopy indicated that to the left of the critical blue line in Figure 1, actin filaments coexist with bundles. To the right of this critical line, the entire network is incorporated into bundles. This difference in $\Gamma_{6k}$ and $\Gamma_{20k}$ states is supported by the DLS analysis (cf. Fig. 3). The fitted relaxation times for each condition are significantly different, and match the states observed in confocal microscopy. Namely, as the molecular weight of the PEG polymers increases, the relaxation time increases because a greater fraction of the network is bundled. The stretched exponent also decreases, indicating a greater heterogeneity in the distribution of bundle sizes. The $\Gamma_{20k}$ state, by nature of being forced into the bundled regime is most likely to be kinetically arrested (47). What would have been free diffusion is limited by inter-bundle steric interactions. In contrast, the control state (no PEG) is most freely diffusing; $\tau_f$ is small and the intensity pattern decorrelates quickly. Since the $\Gamma_{6k}$ morphology exists as a mixture of the control state and the $\Gamma_{20k}$ state it combines features of both the kinetic arrest and the free diffusion modes. Given this, one would expect that the stretched exponent would be smaller for the $\Gamma_{6k}$ state. However, the opposite is observed – where the stretched exponent is smallest for the $\Gamma_{20k}$ state. The reasoning for this is not clear within the context of this analysis. Further quantitative work with our scattering data is difficult when using Mie scattering assumptions, such as spherical particle shapes, which cannot be applied to actin filaments. Studies have shown that dynamic light scattering data can also be analyzed by changing the dynamic structure factor of the fit to a model suited for semiflexible polymers (35–37). The decay rate associated with this semiflexible polymer dynamic structure factor depends on the persistence length of the polymer, which can be used as a fit parameter. With this type of analysis, we report values for the persistence lengths of the $\Gamma_{6k}$ state to the $\Gamma_{20k}$ state, which are statistically indistinguishable from each other yet show increases in respective mean values.

FRET analysis gives the intra-bundle spacing between actin filaments. With PEG as a bundling agent, we expect the filament bundles to be in the fully coupled regime (48). Our analysis gives us a measure on the distance associated with fully coupled actin filaments bundled by depletion forces. This could be extended to measure intra-bundle distance for different actin bundling proteins. The results of the FRET analysis also enable us to estimate the number of filaments incorporated into the bundles given the measured cross-sectional size of bundles from the electron microscopy experiments, the measured intra-bundle distance, and the diameter of an actin filament (44). If we assume a hexagonal packing geometry, we get a measured number of filaments of $N_{FRET} = 217$. It is worth mentioning that our protocol for SEM imaging requires desiccation of the actin network, which may lead to a coalescence of bundles and therefore an overestimation of the bundle thickness. Bundles incorporating 100s of filaments have been observed in drosophila bristles in the presence of different crosslinking proteins, forked proteins and fascin (49–51). In comparison, while our reported values of persistence length via DLS are similar to earlier work with fascin crosslinks (52), results from the semiflexible DLS analysis indicate only less than ten filaments incorporated into bundles for $\Gamma_{6k}$ and $\Gamma_{20k}$ sample conditions. This is determined by calculating the bending modulus from the persistence length (cf. Fig. 4) and applying the known scaling for fully coupled bundles (48). In some part, this can be related to the coexistence of filaments and bundles in the $\Gamma_{6k}$ case. The underestimate is less clear in the $\Gamma_{20k}$ case. In the semiflexible polymer DLS analysis we calibrate one value of the hydrodynamic radius $a$ as a tuning parameter on the control (no PEG) system. In doing this, there isn't any change in $a$ for the $\Gamma_{6k}$ and $\Gamma_{20k}$ fits. Further, with increasing the molecular weight of PEG solutions in water, the viscosity is shown to increase slightly (53). We use one value of kinematic viscosity for all fits which leads to an additional source of error (54). Together, the results from the semiflexible polymer DLS and electron microscopy give bounds on the number of filaments that are incorporated into bundles for the $\Gamma_{6k}$ and $\Gamma_{20k}$ sample conditions.

The physics that limits the extent of bundle diameter is still very much an open question. When well-mixed, we observe that actin filaments bundle in a polydisperse way (cf. Fig. 5e), with bundles forming simultaneously across the full spatial extent of the system. Given this, there are two different mechanisms to consider. The first is the mechanism by which a free actin filament gets incorporated into a neighboring bundle. The other is the interaction between two mature bundles. For the formation of a single bundle theoretical arguments have been proposed in terms of chirality(55), packing defects(56), counterion repulsion (57). Per the bundle-bundle interaction arguments have been presented pointing to the interplay between surface-surface interactions and macroscopic hydrodynamic forces in the system (26). The gels in our study exhibit the polydisperse incorporation of filaments into bundles across the full spatial extent of the system, both in the model and experimentally. The polydisperse morphology of bundles seen in the model only includes surface-surface interactions, suggesting that the hydrodynamic forces aren't as dominant. Future work could investigate the limitations on bundle diameter more systematically via similar measurement techniques.

One question pertains to the nature of the transition between strongly bundled and weakly bundled regimes. It is well known that a system of colloidal hard rods undergoes a first-order phase transition between isotropic and nematic phases at a critical density of rods (58). Introducing depletion interactions produces an additional coexistence phase of isotropic and nematic phases (59). The extent to which actin filaments behave like stiff colloidal rods is not well understood. Actin filaments are semiflexible, with a persistence length of ~10 μm(38), and possess a net negative surface charge (60). Solutions of actin filaments do undergo an isotropic-nematic phase transition, but only upon a critical concentration of approximately 2 mg/mL(61) (cf. 0.5 mg/mL in this work). Isotropic-nematic coexistence phases in actin networks have been observed, as so called nematic tactoids which occur when actin filaments are shorter than 2 μm on average (62). This biphasic behavior is indicative of a first-order phase transition. Introducing depletion interactions in an actin filament network generally results in networks of bundles(25,26,63) as well as higher-order aggregates (63). So far, actin networks experiencing a depletion interaction have been presented as undergoing a transition between single-filament and bundled phases (25). Our work demonstrates that there is an additional coexistence phase containing both single filaments as well as isolated bundles, which we call the weakly bundled phase. (cf. Fig. 1). The transition between the coexistence phase and the strongly bundled phase (demonstrated as a blue line in Figure 1) likely corresponds to a first order phase transition, in analogy to the first-order nature of the isotropic-nematic phase transition. One interesting consequence of the coexistence nature of the weakly bundled regime is the possibility of studying nucleation-growth and spinodal decomposition, which could be characterized by temporally tracking the dynamical evolution of bundles, either by microscopy or light scattering. The effect of depletant molecular weight has been more thoroughly investigated for colloidal particles(64), but less so for actin-based systems. One outcome of our study is that we characterized the boundary between weakly and strongly bundled regimes in terms of PEG molecular weight.

Previous work from Hosek and Tang (26) predicted that the critical concentration of bundling depends on PEG molecular weight and presented experimental results to validate their claim. Our current study expands on this work by exploring the system phases below the critical concentration, namely the weakly-bundled regimes to give a more complete characterization of the phase space for these networks. Furthermore, we provide a more thorough description of the nature of this transition by conducting multimodal material characterization of network states $\Gamma_{6k}$ and $\Gamma_{20k}$.

Further, we corroborate our findings with a phenomenological simulation framework that captures a transition from weakly bundled to strongly bundled states by systematically varying PEG molecular weight and concentration. In particular, we found that the morphology of the network became increasingly bundled along both axes, in agreement with our experimental results (cf. Fig. 2). However, we find one difference between simulation and experiment, namely the nature of the transition between single filament, weakly bundled, and strongly bundled regimes. In the model, we observe a smooth transition in the degree of bundling, whereas in experiment the transition is much sharper. This is presumably a consequence of finite size effects; while the simulation study consists of 200 actin filaments, the experiments study systems have ~ 1000 actin filaments. Furthermore, for computational simplicity, the actin filaments in the simulations were assumed to be about an order of magnitude shorter than in experiments. This effectively reduced the separation between the actin and PEG length-scales in the simulations compared to the experiments, potentially making the transition in simulations less sharp compared to the experiments. The model phase space is built from a choice of model specific interaction parameters (see Methods). These parameters were tuned to recover the same morphological behavior as observed in experiment. In future work, these parameters could be systematically tuned to investigate the effects of changing model interaction strengths which would likely recover the sharpness in phase transition observed experimentally. Further, we could leverage this framework to test for reversibility (26) in network morphology to help motivate studies into reversible soft matter materials.

Multimodal characterization of actin networks driven to bundled states via varying PEG molecular weight and PEG concentration gives distinct insights into a set of material properties for the system. We experimentally observe that there is a transition between a weakly bundled and strongly bundled state that occurs for fixed PEG concentration at 0.1 % w/vol. This type of transition from a weakly bundled to strongly bundled state is also observed in our simulation results. The storage modulus for the $\Gamma_{20k}$ state is larger than for the $\Gamma_{6k}$ state across all replicates, indicating a difference in gel stiffness. The state has a longer diffusion lifetime as shown by dynamic light scattering measurements, perhaps due to the increased stiffness of the bulk material. We also note that we observe stronger gels than previously reported (25). We suspect this is due to in part to the omission of gelsolin in the current work which is known to truncate the length of actin filaments and could impact bulk network properties as a result. Further, the current study uses 50mM KCl vs 1M KCl. Using higher concentrations could impact filament-filament interactions and make the gel weaker. On the scale of an individual bundle, electron microscopy and FRET measurements give bounds on the number of actin filaments incorporated into a bundle for states and further, these modalities inform that bundle diameter and intra-bundle spacing are equivalent across these morphological states. FRET also gives insight into the notion of "effective" crosslinking by parameterizing how close filaments within a bundle are to one another. One might expect to uncover physical signatures between different mechanisms that drive bundling based on extending the FRET methods outlined herein.

## Conclusion

The main takeaway is that actin morphology can be changed at constant actin concentration and constant PEG concentration, by just changing the molecular weight of the PEG molecules. While previous work has predicted this variation(26), this study experimentally demonstrates the existence of this transition and rigorously characterizes the properties of these networks across the transition. The ability to dynamically change actin morphology without adding or removing material from the solvent is similar to a number of mechanisms already found *in-vivo*, such as within condensates(65) or macromolecular crowding (66). The identification of this phase transition offers insight to explain the dynamic properties of cell-based actin manipulation, and perhaps a pathway towards harnessing this protein in artificial contexts. Our work provides a more complete characterization of the depletion interaction, which is necessary to better understand the intracellular organization of the cytoskeleton (18). Furthermore, this research could lead to the development of active rheological modifiers, materials which can change their rheological properties upon application of a fueled stimulus such as light.

## Methods

**Actin Preparation.** Actin was purified from rabbit psoas skeletal muscle from Pel-Freeze using a GE Superdex 200 Increase HiScale 16/40 column and stored at −80 °C in G-Buffer (2 mM tris-hydrochloride pH 8.0, 0.2 mM disodium adenosine triphosphate, 0.2 mM calcium chloride, 0.2 mM dithiothreitol). All protein stocks were clarified of aggregated proteins at 100 000g for five minutes upon thawing and used within seven days. The G-actin concentration in the supernatant was determined by measuring the solution absorbance at 290 nm with a Nanodrop 2000 (ThermoScientific, Wilmington, DE, USA) and using extinction coefficients of 26 600 $M^{-1}$ $cm^{-1}$.

**Confocal Microscopy.** Samples were prepared to yield a final buffer concentration of 20 mM imidazole pH 7.4, 50 mM potassium chloride (KCl), 2 mM magnesium chloride ($MgCl_2$) , 1 mM dithiothreitol, 0.1 mM ATP, 1 mM trolox, 2 mM protocatechuic acid, and 0.1 mM protocatechuic 3,4-dioxygenase. The polyethylene glycol, KCl, imidazole, dithiothreitol, and $MgCl_2$ were purchased from Sigma-Aldrich. The adenosine triphosphate and trolox were purchased from Fisher Scientific. The protocatechuic acid was purchased from the HWI Group.

Glass flow cells were prepared by sonication of individual slides in water for 5 minutes, followed by blow-drying with nitrogen. These slides were then placed in a base piranha solution of five parts DI Water, one part 30% hydrogen peroxide, one part 30% ammonium hydroxide for thirty minutes at 80 °C. These slides were then sonicated again for 5 minutes, blow-dried with nitrogen, and stored in isopropanol until use. Thick coverslips and thin slides were attached by means of melting Parafilm with pre-cut chambers, which are then treated with potassium hydroxide for ten minutes to activate hydroxyl groups on the glass surface and then passivated with poly-l-lysine–g—polyethylene-glycol from Nanosoft Polymers.

We used an Olympus FV1000 motorized inverted IX81 microscope suite, with instrument computer running FV10-ASW software version 4.2b software, to image actin networks using laser-scanning confocal microscopy. Actin filaments were labelled with rhodamine–phalloidin on a one-to-one molar ratio and excited with 543 nm wavelength laser light. Each sample of actin was prepared once and imaged in three random, well-separated locations.

Each z-stack taken was processed using ImageJ. The image was opened and a maximum z-projection across 21 μm through the bulk of the image was produced.

For each confocal z-stack both the degree of bundling and the mesh size were algorithmically determined. The procedures to determine each are as follows. Z-stacks are normalized and subsequently binarized with an Otsu filter. Binarized volumes are then skeletonized using Scikit-Image's skeletonize routine. The number of nodes is then extracted from the skeleton. The number of nodes in the skeleton are taken as proxy for how effective the PEG molecules are in bundling the actin network. For a filamentous network, one would expect the number of skeleton nodes to be large – as one would need more skeletons to describe all unique contours in the actin network. Conversely, for a strongly bundled network, one would expect the number of skeleton nodes to be small. We then take the inverse of the number of nodes to give us our metric for the degree of bundling. For the mesh size calculations, we follow previously developed methods (63,67).

**Rheology.** Experiments were performed on a TA Discovery HR 20 rheometer fitted with an 8mm stainless steel parallel plate geometry. Master buffer was prepared according to the steps described in the Confocal Microscopy section, else the PCA, PCD, Trolox. Master buffer was mixed by pipette with deionized water and PEG molecules of appropriate molecular weight and in the desired concentration. Actin was then added and mixed gently by pipette before adding to a 1:1 molar ratio, actin:phalloidin. After combining and mixing with dried phalloidin, the sample was pipetted onto the rheometer with a total volume of 50 μL. Throughout all experiments, a temperature-controlled Peltier plate maintained the temperature at 25°C, and a solvent trap was utilized to prevent evaporation during data acquisition. Immediately after loading the sample, a time sweep was started to monitor the evolution of the shear moduli over time. The time sweeps were performed at a frequency of 1 rad/s and 1% strain.

**Dynamic Light Scattering.** Actin and master buffer were prepared according to the steps described in the Confocal Microscopy section, else the PCA, PCD, Trolox. Phalloidin in methanol was dried using compressed nitrogen gas and added to the sample solution in a 1:1 molar ratio with the actin to stabilize the filaments and prevent depolymerization. 45 μL were imaged in a Malvern Zetasizer Nano ZS instrument for five runs for each sample. The refractive index of the master buffer was found to be 1.334, and the refractive index of actin was found to be 1.3343.

Particles in a fluid are known to scatter incident light. The diffusion of the suspended particles changes the reported intensity at each angle measured; therefore, using a directed laser, it is possible to use the intensity measurement across multiple temporal decades to understand the Brownian dynamics of the particles in solution. This is done by measuring the diffraction field at a given time $\tau$, and then re-measuring the sample at some later time (68).

$$G_2(\tau) - 1 = G_1(\tau) = \sigma^2 \left( e^{-\left(\frac{\tau}{\tau_f}\right)^\beta} \right)^2 \tag{3}$$

For spherical particles, this is usually accomplished using the Stokes-Einstein equation, which directly gives the hydrodynamics radius. However, for non-spherical particles this calculation is no longer valid, and a quantitative comparison requires a calculation of the stretched exponent: β. Stretched exponent and relaxation time calculations have previously characterized formation of colloidal gels (33) and precursors to contraction of active networks (69).

In the regime of semiflexible polymers, the dynamic structure factor is given by,

$$G_2(\tau) - 1 = G_1(\tau) = G_1(0)\left[ -\frac{\Gamma\left(\frac{1}{4}\right)}{3\pi}\left[\frac{k_B T}{4\pi\eta}\left(\frac{5}{6} - ln(q2a)\right)\right]^{3/4}\frac{q^2 t^{3/4}}{L_p^{1/4}}\right], \tag{4}$$

where $\Gamma$ is the gamma function, $k_B$ is the Boltzmann constant, $T$ is temperature, $\eta$ is the kinematic viscosity, $q$ is the scattering vector, $a$ is the mean hydrodynamic radius of the meshwork and $L_p$ is the persistence length. To perform the fits, the hydrodynamic radius of the system was first set by tuning the persistence length of the control to match the expected value for actin meshwork (38). Subsequently, DLS curves were fit using Eq. 4 directly to extract the persistence lengths.

**FRET.** G-actin, Atto 488, and AlexaFluor 555 maleimide dye were thawed and 1 mL G-buffer 1x was prepared and protein stock was clarified of aggregates as described above. The volume of dye needed to achieve an excess of 10x dye to protein was calculated. The dye was added to G-actin solution and mixed thoroughly by pipette prior to 2 hours of incubation at room temperature, allowing the dye to bind actin cystines. During incubation, Princeton 20 centrispin columns are hydrated with 650uL of G-buffer, allowing the resin to swell for 30 minutes. After incubation, the hydrated columns were placed in a centrifuge and spun at 700g for 2 minutes to remove excess G-buffer. The dyed protein solution was then added to the hydrated column by pipette, being careful to avoid the edges of the column while pipetting. The columns were then placed in a centrifuge and spun at 700g for 2 minutes, where the gel filtration column effectively separates the actin from the free dye. Concentrations were then measured via nanodrop. Labelled G-actin was aliquoted, flash-frozen with liquid nitrogen, and stored at -80℃ until experimentation.

Sample chambers are constructed with piranha etched #1.5 glass microscope coverslips and 2mm thick silicon gaskets with a 5mm diameter hole punched out. Prior to assembly the slides and gaskets are treated in Hellmanex solution at 80℃ for 20 minutes to ensure adhesion between the two components. 30 minutes prior to addition of protein samples, the chambers are passivated with 1 mg/ml bovine serum albumate (BSA) to prevent interactions between the actin and the glass slide. During slide passivation the sample is mixed via a multi-step procedure. The first step is preparing donor seed filaments where half of the total sample, deionized water, and master buffer are mixed with the volume of donor-labelled (Atto 488) actin monomers by pipette. This polymerizes the donor filaments, which are then stabilized by adding to 1:1 phalloidin. The donor seed filaments are then added to the other half of the total sample, deionized water, and master buffer along with PEG molecules, acceptor-labelled (AlexaFluor 555) actin, and unlabeled actin monomers and mixed gently by pipette. This is then added to 1:1 phalloidin to stabilize the filaments. The BSA in the sample well is removed by washing 5 times with master buffer. After the final wash, the master buffer is removed and 20 µL of combined protein sample is added to the sample well. The final concentration for all samples is as follows. [Atto 488 actin] = 0.1µM, [AlexaFluor 555 actin] = 1.0 µM, [Total Actin] = 12 µM.

The sample was loaded onto a home-built time-correlated single photon counting (TCSPC) confocal fluorescence microscope. The microscope utilizes a 486nm picosecond laser with a 50MHz repetition rate and laser power set to 50µW via neutral density attenuation. For all samples, the laser focus was place at a depth of 5 microns from the base of the microscope slide using a Mad City Laboratories piezo stage and translated via micrometer adjustment in x and y dimensions to navigate to different spatial locations. Emitted photons were collected using a 1.45 NA, 100x magnification microscope objective and routed through a pinhole and 511/10 bandpass emission filter toward a Hamamatsu GaAsP photomultiplier tube. Photomultiplier output pulses were then amplified and counted with a Becker and Hickl (BH) TCSPC computer card. For each replicate, 3 random, well-separated regions were imaged to incorporate intra-sample variation in our results.

Lifetime data was analyzed using BH SPCImage software where decay-matrix calculations were performed to generate a distribution of fluorescence lifetimes within an image. The quantity $\tau_{DA}$ is determined from this distribution in fluorescence lifetimes. The quantity $\tau_D$ was determined with a control study where the distribution of fluorescence lifetimes was measured for actin filaments that were only labelled with the donor. The average donor-to-acceptor distance was then calculated using equations (1) and (2).

**Electron Microscopy.** Experiments were performed on a Thermo Scientific Scios 2 DualBeam through the Texas Materials Institute (TMI). Sample holders were constructed using silicon wafers purchased from Montco Silicon Technologies Inc. The wafers were cut with a diamond scribe to match the geometry of the EM sample stubs. The silicon wafers were cleaned using a rinse of deionized water, followed by blow drying with a clean nitrogen line. The scribed silicon substrates were adhered to the sample stubs using double-sided carbon dots adhesive stickers. The silicon wafers were connected to the sample stub via aluminum tape to ensure proper electrical contact through the sample.

Master buffer was prepared according to the steps described in the Confocal Microscopy section, else the PCA, PCD, Trolox. Master buffer was mixed by pipette with deionized water and PEG molecules of appropriate molecular weight and in the desired concentration. Actin was then added and mixed gently by pipette. The sample was then added to a 1:1 molar ratio, actin:phalloidin and subsequently mixed. The sample was then pipetted onto the silicon wafer in a manner such as to ensure a flat film of solution across the surface as much as possible.

Samples were transported to TMI and loaded into an EMS sputter coater. The sputter coater was run with the following settings: Current = 40mA, Deposition time: 45 seconds, Species = Au/Pt: 60/40. The sputter coated sample was then transferred directly to the Scios 2 Dual Beam for imaging. For Scanning Electron Microscope (SEM) depositions the following settings were used: Carbon deposition – 15x10x0.05 µm at 5kV, 3.2nA. For SEM imaging the following settings were used: OptiPlan Mode 2kV, 1.2nA, Working Distance 7mm. FIB processing was carried out as follows: Step 1: Pt deposition 15x10x1µm at 30kV, 1nA; Step 2: Regular Cleaning Cross-section 17x10x10µm at 30kV, 15nA; Step 3: Cleaning Cross-section 18x5x12µm at 30kV, 1nA. Stage tilt was at 52 degrees for all cross-sectional cuts.

Image analysis was performed using Python where an active contours method was implemented to measure critical dimensions of actin bundles as visualized by SEM on fresh cross-sectional FIB cut surfaces.

**Simulation Parameters.** The actin bead-chains and have a uniform length of 160nm and bead diameter ($\sigma$) of 8nm, and the diameter of the PEG spheres is varied from 0.25 $\sigma$ to 0.75 $\sigma$. All distances in the simulations are scaled by $\sigma$, and all times by the time ($\tau$) it takes for a bead in an actin bead-chain to diffuse across a distance $\sigma$. The diffusion coefficient for a freely diffusing actin monomer is estimated to be $10^1 \mu m^2 s^{-1}$, which gives $\tau = 0.1 s$ (70). All energies in the model system are scaled by $K_b T$. The simulation box size is set to $75\sigma$ by $75\sigma$ by $10\sigma$ across all simulations, and periodic boundary conditions are enforced.

To understand how the (1) concentration and (2) molecular weight of the PEG particles influenced the actin bundling, we vary (1) the number of PEG particles from 4000 to 10000 while keeping the simulation box volume unchanged and (2) PEG particle radius from 2nm to 6nm, respectively. To convert from PEG radius of gyration (Rg) to molecular weight (M), we assumed the classic result from Flory theory, $R_g \propto M^{\frac{3}{5}}$. This gives molecular weights of 6k, 20k, and 37.4 k for radii of $0.25\sigma$, $0.5\sigma$, and $0.75\sigma$ respectively. The PEG concentration is varied from $4000\frac{V_\rho}{V_{box}}$, $5000\frac{V_\rho}{V_{box}}$, $6000\frac{V_\rho}{V_{box}}$, and $10000\frac{V_\rho}{V_{box}}$, where $V_\rho = \frac{4}{3}\pi r^3$, $r$ representing the effective radius of and $V_{box}$ is the volume of the simulation box which is $56250 \ \sigma^3$. We explore 12 different systems to determine the effects of PEG concentration and effective molecular weight. The number of actin filaments is held constant at 200 for all 12 systems.

For the Lennard Jones pairwise interactions, $\epsilon_{PEG-PEG}$ is set to 1.25 for attractive interactions and 1.0 for repulsive interactions; $\epsilon_{actin-actin}$ is set to 0.3 for attractive interactions and 1.0 for repulsive interactions. We encode only repulsive interactions between the PEG beads and the actin strands, and $\epsilon_{PEG-actin}$ is equal to 2.0.

## Author Contributions

## Conflicts of interest

There are no conflicts to declare.

## Acknowledgements

This research was primarily supported by the National Science Foundation through the Center for Dynamics and Control of Materials: an NSF MRSEC under Cooperative Agreement No. DMR-1720595 with additional support from the Welch Foundation (F-1848 and F-1696). The authors acknowledge the use of facilities and instrumentation supported by the National Science Foundation through the Center for Dynamics and Control of Materials: an NSF MRSEC under Cooperative Agreement No. DMR-1720595. We also acknowledge the Texas Materials Institute for the use of facilities and instrumentation. We would also like to recognize Carlos Baiz, and Xiaobing Chen for insightful discussions.

## Notes and references

# Supplemental Information

The supplemental information includes raw time series data replicates for bulk rheological measurements shown in Figure S1.

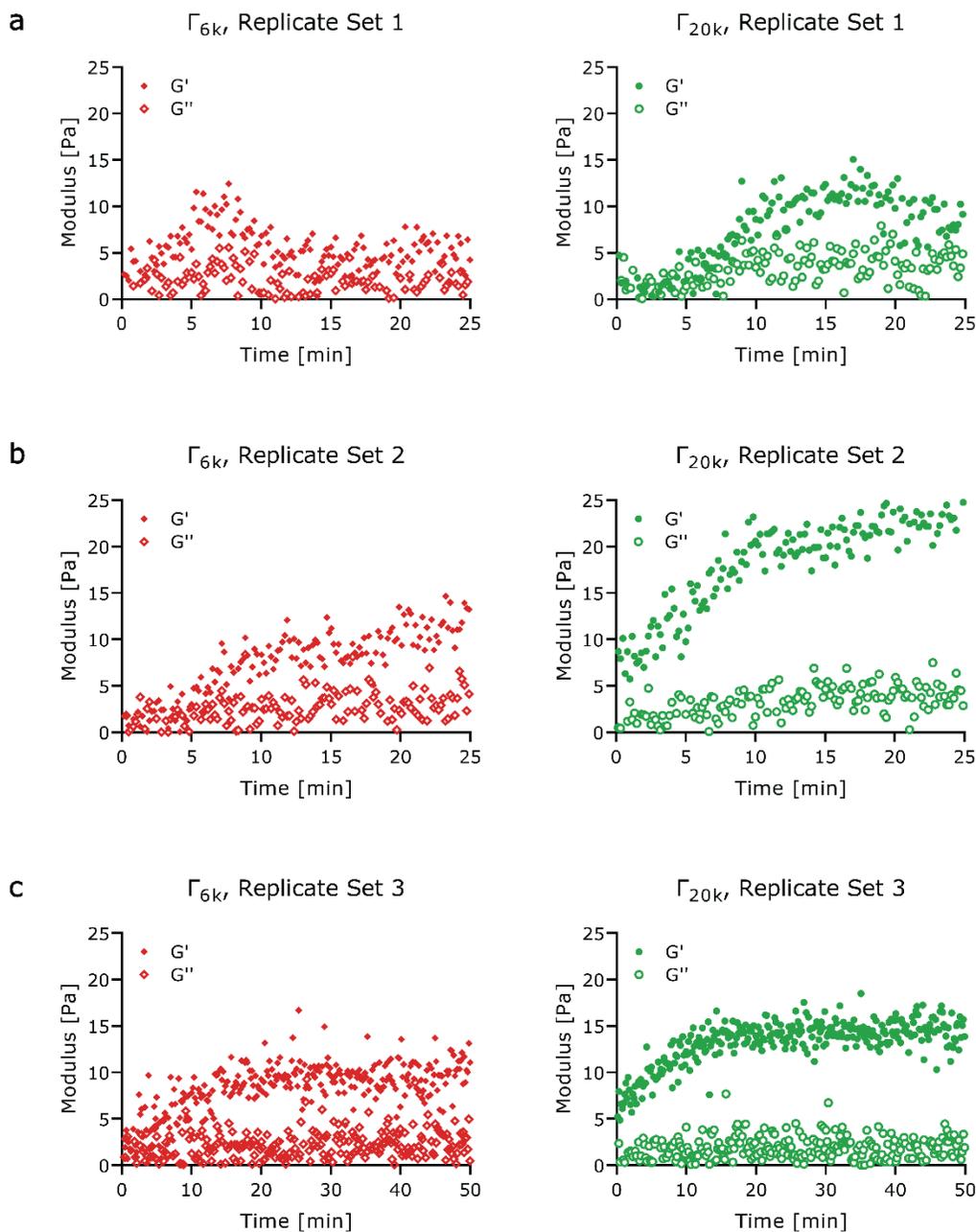

Figure S1: Rheometric measurements (*G'* and *G''*) are reported all pairs of replicates (a, b and c) of $\Gamma_{6k}$ (red) and $\Gamma_{20k}$ (green), formulated and measured with the same parameters as those used for the data in Figure 7. The data is reported as time sweeps showing the evolution of the storage moduli of the bundled actin networks formulated with (red diamonds) and (green circles).